# Influence of the ferroelectric polarization on the electronic structure of BaTiO$_3$ thin films

N. Barrett[1,*], J. Rault[1], I. Krug[2], B. Vilquin[3], G. Niu[3], B. Gautier[4], D. Albertini[4] and O. Renault[5]

[1] *CEA, IRAMIS/LENSIS/SPCSI, Bâtiment 462, 91191 Gif sur Yvette cedex, France*
[2] *Institut f. Festkörperforschung IFF-9, Forschungszentrum Jülich, 52425 Jülich, Germany*
[3] *Institut of Nanoscience de Lyon (INL), 36, avenue Guy de Collongue, 69134 Ecully, France*
[4] *Institut National des Sciences Appliquées de Lyon (INSA), 20, avenue Albert Einstein, 69621 Villeurbanne, France*
[5] *CEA LETI/MINATEC, SPCIO, 17 avenue des Martyrs, 38054 Grenoble cedex, France*

*\* corresponding author: nick.barrett@cea.fr*

**Abstract:** Micrometric domains of precise ferroelectric polarization have been written into a 20 nm thick epitaxial thin film of BaTiO$_3$(001) (BTO) on a Nb doped SrTiO$_3$ (STO) substrate using PiezoForce Microscopy (PFM). The domain dependent electronic structure has been studied using fully energy-filtered PhotoEmission Microscopy (PEEM) and synchrotron radiation. Shifts, induced by ferroelectric polarization, of up to 300 meV are observed in the work function of the sample. The surface is Ba-O terminated. Polarization-induced distortion of the electronic structure is observed in the valence band and on the Ba 3d, Ti 2p and O 1s core levels of BTO. Polarization dependent surface adsorption is observed. A simple electrostatic model based on net surface charge is not sufficient to explain the observed modifications in the electronic levels.

## Introduction

Thin film ferroelectrics are an active field of research due to their promising technological applications in electronic devices such as ferroelectrics memories. Their surfaces also show fascinating properties such as catalysis enhancement or possible magnetoelectric coupling with magnetic overlayers. In a thin film the polarization state gives rise to a fixed surface charge, $\sigma = \vec{P}\cdot\vec{n}$, where $\vec{P}$ is the polarization vector and $\vec{n}$ is the surface normal. [SNGR09] This gives rise to a depolarizing field which reduces the FE polarization in the material. The surface state of thin films is therefore crucial in determining the electrical properties of real thin film based devices. The surface charge may be screened either by internal charge carriers and defects in the ferroelectric or externally by adsorbates. Thus, surveying the polarization dependence of electronic structure is a fundamental step towards the understanding of some very promising phenomena.

Epitaxially strained BaTiO$_3$ (BTO) films show interesting ferroelectric properties compared to those of bulk BTO. The maximum in-plane strain can attain 1-2%. Theory has predicted that the Curie temperature can attain 550°C. Ferroelectric (FE) BTO(001) is an ABO$_3$ perovskite oxide with a tetragonal structure. The ferroelectric distortion lies along the c axis. For example, Fig 1 shows the expected distortion for a Ba-O terminated surface unit cell, with P$^-$ polarisation giving a net negative surface fixed charge, as indicated.

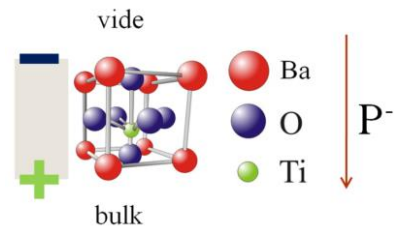

**Figure 1.** Distortion for a Ba-O terminated unit cell

X-ray photoemission microscopy (XPEEM) conserves the spatial origin of the photo-emitted electrons [EWRB09]. Domains with different ferroelectric polarization should show shifts in the positions of the electronic levels. Local photoelectron spectra from the different FE domains will contain this information. XPEEM collects electrons with a small inelastic mean free path, it is thus surface sensitive. It is non destructive, crucial if one wishes to carry out other imaging, structural or electrical analyses of the same sample. In a simple electrostatic model, the net surface charge should modify the kinetic energy of the photoelectrons.

## Experimental

A 20 nm thin film of BTO has been grown epitaxially on an Nb-doped SrTiO$_3$ (001) substrate by Pulsed Laser Deposition (PLD). Thanks to niobium doping, the substrate is conducting and works as an electrode. It also prevents significant charging effects during photoemission. Micron scale polarized ferroelectric domains have been written by Piezo Force

Microscopy (PFM, an Atomic Force Microscope with a conductive tip) in d.c. (writing) mode as shown on the Fig. 2 together with a typical phase signal as observed by PFM in the a.c. (reading) mode. The topography image is uniform. Thus the written region is flat but the phase signal shows FE domains with a clear electrical contrast.

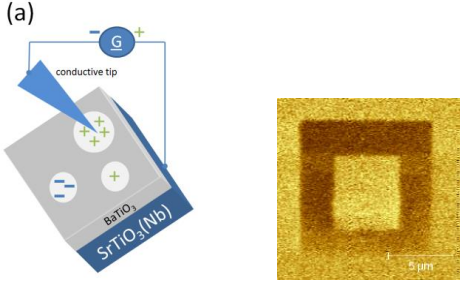

**Figure 2.** (a) schematic of a PFM. (b) PFM phase image of the BTO thin film

The thin film deposition process includes an oxygen plasma treatment at 620°C and a partial pressure of 3 Pa. After plasma treatment, the sample was cooled in $4\times10^4$ Pa oxygen. It is thought that the film has undergone a phase transition to a single ferroelectric domain with a P$^-$ imprint at RT. This is confirmed by the PFM writing which was easier for P$^+$ followed by P$^-$ poling, showing that the film has actually a P$^-$ imprint. Two domains with strong negative polarization, called P$^{--}$ and P$^{---}$, were written by PFM in the film.

The energy filtered XPEEM experiments have been conducted on TEMPO beam line of the SOLEIL synchrotron (Saint Aubin, France). The temporary end station is a NanoESCA XPEEM (Omicron Nanotechnology GmbH) [EWNZ05, RBBZ07]. This aberration-compensated energy-filtered PEEM offers a lateral resolution of ~ 100 nm for core level emission with, simultaneously, an energy resolution of 0.15-0.2 eV [EWRB09]. Experiments were conducted in Ultra High Vacuum (UHV) at a pressure of $2.10^{-8}$ Pa. To avoid a transition to the paraelectric phase, the sample was not heated after introduction into the XPEEM chamber. Unfortunately, some surface contamination is therefore inevitable, and as we will see, may reduce the observed FE contrast.

Image series were acquired over the photoemission threshold region (photon energy hν = 575 eV), the Ba 3d, Ti 2p and O 1s core levels (hν = 898, 575 and 647 eV respectively) and the valence band (hν = 80 eV). The photon energy was tuned with respect to each core level so as to maintain the same surface sensitivity and to avoid Auger lines interfering with the core level spectra (photoelectron kinetic energy between 100-120 eV). A 64.5 μm Field of View (FoV) was used. 12 kV extraction voltage. Dark images were also acquired with the beam shutter closed in order to remove camera noise. The non-isochromaticity of the XPEEM is smaller than 60 meV.

**Results**

*Work function*

With photon energy of 575 eV, the threshold represents a true secondary electron (SE) peak and direct transitions play no part in the intensity position. Fig. 3 shows a selection of images from the total image series. Contrast inversions between differently FE domains as a function of $E - E_F$ are immediately visible. The position of the threshold in the local spectra extracted from the image series is equivalent to the work function. After correction for the Schottky effect due to the high extractor field, ΔE = 98 meV for 12 kV, [RBRH06] the position of the photoemission threshold corresponds exactly to the work function. The threshold spectra were fitted with a complementary error function

$$I(E) = A \cdot erfc\left(\frac{\Phi_0 - E}{\sqrt{2}\sigma}\right) + I_{min}$$

where $\Phi_0$ is the work function and σ the half-width of the rising side of the secondary electron peak. The work function shifts up to 300 meV depending on the domain polarization strength.

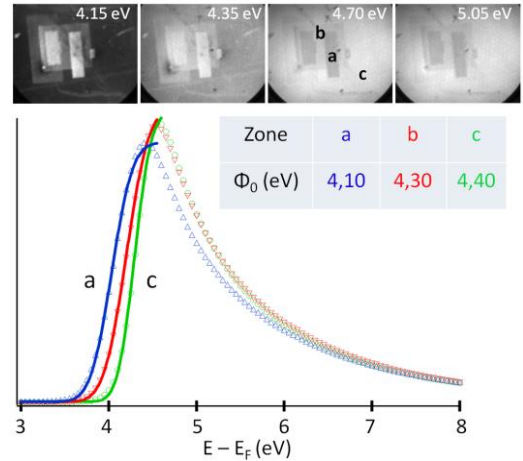

**Figure 3.** (upper) images from the threshold image series as a function of (E – E$_F$); (below) threshold spectra extracted from regions (a), (b) and (c) and the complementary error function fits to the rising side of the SE peak.

*Valence band*

The local valence band spectra extracted from the three differently polarized regions are shown in Fig. 4. The Ba 5p emission is always centred at ~ 15 eV and does not change in intensity as a function of FE polarization. The weak component centred between 9.0 and 9.5 eV increases steadily in intensity from zone (a) to (c). The energy position of the intensity maximum of the valence band shifts slightly to higher binding energies as the strength of the polarization increases, however, the shift is well below the experimental resolution. A zoom on the leading edge of the valence band is shown in the inset. A 200 meV shift between the position in zone (a) and that for (b) and (c) is observed. This translates into a shift in the valence band maximum (VBM), as obtained from a linear extrapolation. A non-zero density of states is always observed in the gap region above the valence band maximum up to the Fermi level ($E_F = 0$ eV in all figures).

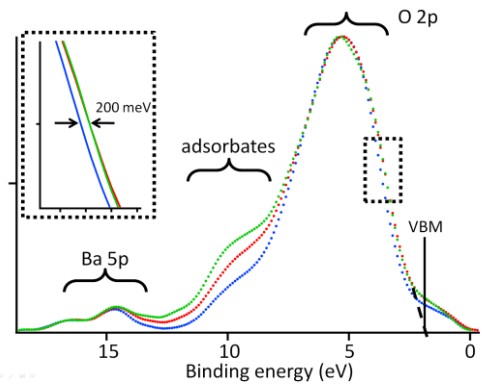

**Figure 4.** Valence band spectra for the three polarized regions ((a) blue; (b) red; (c) green). (Inset) enlargement of the valence band leading edge. The thin black and grey lines are two theoretical DOS calculations, see text.

*Core-levels*

A Shirley background was used. Each core level component was fitted with a Voigt function, which in fact turned out almost Gaussian. The full width half maxima (FWHM) were 1.5, 1.2 and 1.4 eV for Ba, O and Ti core levels, respectively.

Fig. 5 shows the Ba $3d_{5/2}$ peak for the three distinct FE domains. There are two components, one from the bulk-coordinated barium (light grey), and the other from the surface barium (dark grey). The surface component is the strongest, confirming that the photoelectron kinetic energy has been set to minimize the inelastic mean free path (~ 0.5 nm). There is some debate as to the exact origin of the surface peak. It may be due to Ba emission from the Ba-O termination layer, but its strength suggests that it could also be due to a space charge region affecting the measured Ba $3d_{5/2}$ binding energy over several unit cells. [LLLM08] This will be the subject of a more in-depth study. Both surface and bulk like components are shifted towards the Fermi level by 150 meV (b) and 300 meV (a) with respect to the unwritten (c).

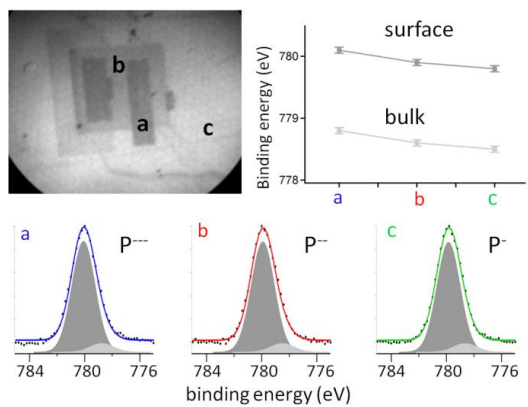

**Figure 5.** (upper left) SE image of the FE domain structure; (upper right) Ba $3d_{5/2}$ core level shifts; (lower) best fits to local spectra extracted from zones a, b and c. Surface components in dark grey, bulk in light grey.

There are three components in the O 1s spectrum, two of which, at higher binding energies (HBE), are again surface related. The first surface component reflects surface, or near-surface oxygen in the BTO, while the second is linked to adsorbed contaminants (most probably, $CO_2$). Fig. 6 shows the O 1s peak (hν = 647 eV) for three different polarizations. The observed shifts are similar to those of the Ba $3d_{5/2}$ core level and the work function.

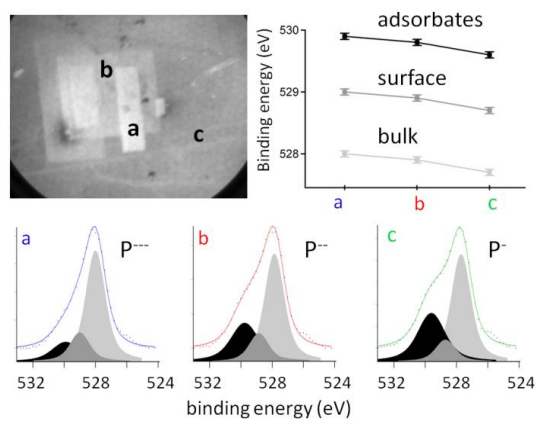

**Figure 6.** (upper left) SE image of the FE domain structure for $(E - E_F) = 4.10$ eV; (upper right) O 1s core level shifts; (lower panels) best fits to local spectrum extracted from zones a, b and c. Surface component are in dark grey, bulk in light grey and contamination in black.

Fig. 7 shows the Ti $2p_{3/2}$ peak for the three domains. The core level shift as a function of polarization is significantly smaller than for the Ba $3d_{5/2}$. This is also an indication that changes in the electronic levels will require an explanation which goes beyond the simple model of a uniform electric field induced by the polarization. The shifts of the Ti $2p_{3/2}$ with respect to the position in zone c are 50 meV (b) and 150 meV (a).

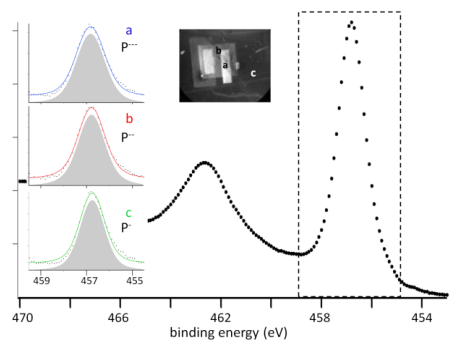

**Figure 7.** Ti 2p spectrum. The central inset shows the threshold image. On the left, the local Ti $2p_{3/2}$ spectra extracted from the three zones a, b and c.

## Discussion

### Work function

A simple explanation for the work function shifts could be the following. For a $P^-$ polarization state, the fixed (negative) surface charge will act to reduce the work function, defined as the vacuum level with respect to the Fermi level. Thus, the more negative the polarization, the smaller the effective work function. Positive polarization, with positive fixed surface charge would have the opposite effect. This electrostatic interpretation is valid for the surface potential and therefore the work function. Since there should be no chemical change in the ferroelectric state, we could assume that the electron affinity remains the same, thus a shift in the work function should be followed by all of the electronic levels. However, as we will see, this is not the case.

### Valence band

The intensity centred at 9 eV in the valence band can be attributed to either $H_2O$ or $CO_2$ adsorption [MSRM88], but not to the formation of a surface carbonate species which would, for example, shift the Ba 5p core levels. Indeed, Baniecki et al [BKY08] show convincing evidence based on DFT calculations for molecular $CO_2$ adsorption. There is clearly more adsorption on the more positively polarized domains.

The sign of the VBM shift follows the electrostatic argument given above. The absence of a Ba 5p core level shift is proof that different electronic levels do not react in the same way to the FE polarization.

### Core levels

The core levels shift systematically to higher binding energies with increasing polarization $P^-$. The Ba 3d shift is twice that of the Ti 2p. The formal ionic charges of Ba and Ti are 2+ and 4+, respectively, but in perovskite oxides there is always some degree of covalency. Several theoretical studies [BMFB96, SI94, EV08] have shown that the effective valence charge on Ti is more strongly reduced than on Ba. The higher ionicity of Ba may explain a bigger FE induced core level shift. However, the notion of valency must be treated with care. Ghosez et al [GMG98] show that dynamical atomic charge provides a better description of variations in bonding hybridization and that charge transfer cannot be treated as a purely local phenomena.

The adsorbate origin of the HBE O 1s component is well documented on STO [MSRM88] and BaO [BKY08]. The local O 1s spectra from our sample indicate that the most polarized zones are the least contaminated. It has been shown that both the adsorption rate and saturation depend on the strength of the surface FE polarization [LZJK08], a higher $P^+$ should increase adsorption, improving surface screening and thus reducing the depolarization field. This is what we observe; the most positive FE domain gives the highest adsorbate signal. A more complete description of adsorption dynamics should include the chamber pressure and also the surface desorption frequently observed when using high brilliance synchrotron radiation. The resultant effective electric field would be a balance between polarization-enhanced adsorption and radiation stimulated desorption.

## Conclusion

We have used energy filtered X-ray photoemission electron microscopy to explore ferroelectric polarization dependent changes in the electronic structure of BTO(001) thin film. $P^-$ domains were poled using Piezo Force Microscopy. Contrast is observed in the work function, core level binding energies and valence band maxima positions. A simple electrostatic model assuming a uniform net electric field, predicts the direction of the observed energy shifts. However, the model of rigid displacement of all the electron bands is not correct. The work function shifts by 300 meV for the most strongly poled domain, followed by the Ba $3d_{5/2}$ core levels. Differences in the Ba and Ti core level shifts suggest that an interpretation including the level of ionicity/covalency is important. The valence band maximum of the most strongly poled domain shifts by 200 meV away from the Fermi level. Polarization dependent screening by adsorbates of the fixed surface charge certainly plays a role in the amplitude of depolarizing field.

## Acknowledgements

We acknowledge SOLEIL for provision of synchrotron radiation facilities and we would like to thank F. Sirotti, the TEMPO beamline staff and B. Delomez for technical assistance.

## References

[BMFB96] A.E. Bocquet, T. Mizokawa, A. Fujimori, S.R. Barman, K. Maiti, D.D. Sarma, Y. Tokura, M. Onoda, *Phys. Rev. B* **1996**, *53*, 1161.

[BKY08] J.D Baniecki, M. Ishii, K. Kurihara, K. Yamanaka, T. Yano, K. Shinozaki, T. Imada, K. Nozaki, N. Kin, *Phys. Rev. B* **2008**, *78*, 195415.

[CBLS04] K.J. Choi, M. Biegalski, Y.L. Li, A. Sharan, J. Schubert, R. Uecker, P. Reiche, Y.B.Chen, X.Q. Pan, V. Gopalan, L.Q. Chen, D.G. Schlom, C.B. Eom, Science **2004**, *306*, 1005.

[CDKG04] S.A. Chambers, T. Broubay, T.C. Kaspar, M. Gutowski, M. van Schilfgaarde, *Surf. Sci.*, **2004**, *554*, 81.

[CKL00] C. Cheng, K. Kunc, and M.H. Lee, *Phys. Rev. B*, **2000**, *62*, 10409.

[EV08] R. I. Eglitis and David Vanderbilt, *Phys. Rev. B*, **2008**, *77*, 195408.

[EWRB09] M. Escher, K. Winkler, O. Renault, and N. Barrett, *J. Elec. Spectros. Relat. Phenom.*, **2009**, doi:10.1016/j.elspec.2009.01.001

[EWNZ05] M. Escher, N. Weber, M. Merkel, C. Ziethen, P. Bernhard, G. Schonhense, S. Schmidt, F. Forster, F. Reinert, B. Kromker, D. Funnemann. *J. Phys. Condens. Mat.*, **2005**, 17, S1329.

[GMG98] P. Ghosez, J.-P. Michenaud, X. Gonze *Phys. Rev. B*, **1998**, *58* 6224.

[LZJK08] D. Li, M. H. Zhao, J. Garra, A. M. Kolpak, A. M. Rappe, D. A. Bonnell, and J. M. Vohs, *Nat. Mat.*, **2008** *7*, 473.

[MSRM88] D. Mueller, A. Shih, E. Roman, T. Madey, R. Jurtz, R. Stockbauer *J. Vac. Sci. Technol. A*, **1988**, *6* 1067.

[PISK03] S. Piskunov, PhD thesis, University of Osnabrück **2003**.

[RBRH06] O. Renault, R. Brochier, A. Roule, P.H. Haumesser, B. Krömker, D. Funnemann, *Surf. Inter. Anal.*, **2006**, *38***,** 375.

[RBBZ07] O. Renault, N. Barrett, A. Bailly, L.F. Zagonel, D. Mariolle, J.C. Cezar, N.B. Brookes, K. Winkler, B. Kroemker, D. Funnemann, *Surf. Sci.*, **2007**, *601*, 4727.

[SI94] C. Sousa and F. Illas, *Phys. Rev. B*, **1994**, *50*, 13974.

[SSH03] H. Salehi, N. Shahtahmasebi, S.M. Hosseini, *Eur. Phys. J. B*, **2003***, 32*, 177.

[ZBRB08] L.F. Zagonel, N. Barrett, O. Renault, A. Bailly, M. Baeurer, M. Hoffmann, S.-J. Shih, D. Cockayne, *Surf. Inter. Anal.*, **2008**, *40*, 1709.

[LLLM08] X. L. Li, H. B. Lu, M. Li, Z. Mai, H. Kim, Q.J. Jia, *Appl. Phys. Lett.*, **2008**, *92***,** 012902.

[SNGR09] J. Shin, V.B. Nascimento, G. Geneste, J. Rundgren, E.W. Plummer, B. Dkhil, S.V. Kalinin, A.P. Baddorf *Nano Letters*, **2009**, *9* 3720.